\documentclass[%
reprint,
amsmath,amssymb,
aps,
pra,
]{revtex4-2}

\usepackage{graphicx}
\usepackage{dcolumn}
\usepackage{bm}

\begin{document}
	
\preprint{APS/123-QED}

\title{Probing short-range gravity using quantum reflection}

\author{J. Boynewicz}
\affiliation{%
	Department of Physics, The University of Texas at Austin, Austin, 78712, Texas, USA
}%
\author{C. A. Sackett}
\affiliation{%
	Department of Physics, University of Virginia,
	Charlottesville, Virginia 22904, USA
}%

\date{\today}

\begin{abstract}
Quantum reflection occurs when ultra-cold atoms are incident on a material surface with sufficiently low
velocity. The reflecting matter wave can interfere with the incident wave to form a detectable pattern, and this
pattern 
contains information about atom-surface interactions at micrometer scales. We discuss how such an interferometer
could be used to probe for anomalous short-range forces that are predicted by some beyond-standard model
theories. We compare a simple analytical model for the anomalous phase
to numerical solution of both the Schr\"odinger and Gross-Pitaevskii equations, finding good agreement.
With interactions, the phase does depend on the atomic density, which can be a source of noise. We nonetheless
predict that under realistic conditions, the reflection technique can reach sensitivities approaching
those obtained with macroscopic objects, and significantly improve the limits on anomalous coupling to atoms.
\end{abstract}

\maketitle

\section{Introduction}

Many models for physics beyond the standard model lead to predictions of new short-range
forces near material objects \cite{ArkaniHamed1999,Adelberger2003}, 
including
forces from axion \cite{Moody1984,Frieman1995,Hall2005b} 
and chameleon \cite{Khoury2004,Mota2006} fields. 
The significance of these predictions has motivated
experimental tests, the most sensitive of which measure the force between two macroscopic objects \cite{Long2003,Geraci2008,Chen2016,Lee2020,Tan2020}.
However, measurements using microscopic objects like single atoms are also important, as some
theories predict a suppression of the force on macroscopic objects. The suppression could arise from
spin-dependent effects \cite{Antoniadis2011,OHare2020,JacksonKimball2023}, 
as expected for axions, or from the chameleon field mechanism \cite{Burrage2015}. 
Experiments
probing the force between neutral atoms and macroscopic objects have not revealed any anomalies
\cite{Hamilton2015,Schloegel2016,Jaffe2017,Panda2024},
but comparatively little data is available about such forces at sub-mm length scales
\cite{Harber2005}.

We propose here a method to search for gravity-like forces between a macroscopic object and 
an atom, based on the interference of matter waves produced by quantum reflection. Quantum reflection
is a universal process that occurs when a sufficiently slow atom approaches a material surface
\cite{Shimizu2001,Mody2001,Pasquini2004,Pasquini2006}. 
If the de Broglie wavelength of the atom is large compared to the range of the attractive Casimir-Polder
surface interaction, then the atom experiences the surface interaction as a sharp step from
which it will reflect, much as an atom reflects from an attractive square-well potential
in elementary quantum mechanics. Any novel force would be small compared to the
electromagnetic Casimir-Polder force, so it would not have an appreciable effect on the reflection
amplitude. Instead, we propose here to observe the quantum phase shift imparted to an atom during the
reflection process via an interferometric measurement.
Notably, this approach does not require a high reflection probability, which
reduces the impact of surface imperfections and characteristics.
We find that at distance scales near 10~\textmu m, we obtain an atomic-scale sensitivity
that significantly improves on existing bounds,
and approaches the sensitivity achieved using macroscopic objects.

A different method for probing atom-surface interactions was recently proposed by Bennett 
and O'Dell \cite{Bennett2019},
in which cold atoms are confined in an optical lattice near the surface. The force between
the atom and the surface is probed by measuring the local Bloch oscillation frequency.
Bloch oscillation have been measured with high precision for atoms in free space,
but achieving similar precision for atoms very near a surface may be experimentally challenging.

Both the Bennett technique and the method proposed here rely on Casimir-Polder shielding
to distinguish the small anomalous force from the large electromagnetic interaction. The shield consists
of a thin layer of conducting material that is placed between the atom and the test mass being probed.
This suppresses the electromagnetic interaction with the test mass, so any changes produced
by the test mass can be attributed to an anomalous force. 
Here we consider
a fixed conducting membrane in front of a movable test mass structure.

In Section II of this paper, we present a simple analytical approximation for the quantum reflection
phase shift. In Section III, we develop a more realistic model based on numerical solution
of the Schr\"odinger and Gross-Pitaevskii equations, 
and find that agreement with the analytical model is generally
good. In Section IV, we discuss how an experimental measurement could be set up and estimate
the sensitivity that could be achieved. Finally, Section V discusses possible extensions
of the approach, along with a summary and conclusions.

\section{Analytical Model}

We first consider an approximate analytical model for the quantum reflection process.
The interaction between an alkali-metal atom and a surface can be described by the Casimir-Polder potential \cite{Casimir1948}
\begin{equation} \label{betaeqn}
	U(x) = -\frac{C_4}{(x+3\lambda_a/2\pi^2)x^3} \equiv 
	- \frac{\hbar^2 \beta^2}{2m(x+3\lambda_a/2\pi^2)x^3}
\end{equation}
where $x$ is the distance to the surface, $\lambda_a$ is the wavelength of the principal
transition of the alkali, and $C_4$ or $\beta$
are alternative parameters to express the
atom-surface interaction strength. For now, we consider $x \gg \lambda_a$ and approximate
$U \propto x^{-4}$. We assume that any surface roughness is small compared to the relevant
length scale of a few micrometers, and that any variations in the surface potential are
slow compared to the time required for an experimental measurement.

Although quantum reflection is a distributed effect, the effect is
strongest at the position where $U(x)$ changes the most quickly relative to the local de Broglie
wavelength of the particle \cite{Cote1997,Carraro1998,Shimizu2001}. 
This position can be determined by maximizing the quantity
\begin{equation} \label{Ucondition}
	\left| \frac{1}{k(x) U} \frac{dU}{dx}\right|,
\end{equation}
where $k(x)$ is the local wave number
\begin{equation}
	k(x) =\left[\frac{2m}{\hbar^2}\Big(E_0 - U(x)\Big)\right]^{1/2} 
\end{equation}
for a particle with incident energy $E_0$. Setting $E_0 = \hbar^2 k_0^2/2m$ leads to 
\begin{equation} \label{kofx}
	k(x) = \left(k_0^2 + \frac{\beta^2}{x^4}\right)^{1/2},
\end{equation}
and \eqref{Ucondition} is maximized at position
\begin{equation} \label{xm}
	x_0 = \left(\frac{\beta}{k_0}\right)^{1/2}.
\end{equation}
We interpret $x_0$ as the point of closest approach for the reflecting particle in
a semi-classical model. 

The quantum reflection process is efficient only for incident velocities less than
$v_c = \hbar/(4m\beta)$. As noted above, high reflection is not
necessary for an interferometric measurement since the amplitude of the detected signal
will scale as the square root of the reflection probability. Pasquini {\em et al.}\ observed reflection probabilities
above $10^{-2}$ for incident velocities up to $5v_c$ \cite{Pasquini2004}.

We are interested in how the phase of the reflected quantum wave 
changes in the presence of an anomalous short-ranged interaction. We suppose the
interaction between point masses $m$ and $M$ separated by distance $r$ 
to have the Yukawa form \cite{Floratos1999}
\begin{equation}
	V(r) = -\frac{GMm}{r} \alpha e^{-r/\lambda},
\end{equation}
where $G$ is Newton's constant, $\alpha$ sets the strength of the interaction relative to gravity, and $\lambda$ sets the range. The potential can be integrated to determine the net interaction
between a particle and an infinite half-plane with mass density $\rho$, yielding
\begin{equation} \label{yukawa}
	V(x) = -2\pi G \rho m \alpha \lambda^2 e^{-x/\lambda}.
\end{equation}

To estimate the phase shift produced by the Yukawa interaction, we use the semiclassical
result \cite{Cronin2009}
\begin{equation} \label{phaseintegral}
	\phi = -\frac{1}{\hbar} \int V(x)\,dt,
\end{equation}
where the integral is over the classical path taken by the particle in the Casimir-Polder
potential. We assume that the particle is incident from infinity with 
speed $v_0 = \hbar k_0/m$, 
approaches to distance $x_0$ from \eqref{xm}, 
and then reflects and returns to infinity. The exact
classical motion cannot be expressed in closed form, but we observe that
at position $x_0$, the local wave number is $\sqrt{2} k_0$, corresponding to classical velocity
$\sqrt{2} v_0$. This is not very different from $v_0$, so we make the simple
approximation that the particle moves at constant velocity throughout the reflection.
Then in Eq.~\eqref{phaseintegral}, we use $dt = \pm dx/v_0$ to obtain
\begin{equation}
	\phi = -\frac{2}{\hbar v_0} \int_{x_0}^{\infty} V(x)\,dx
		= \frac{4\pi G\rho m}{\hbar v_0} \alpha \lambda^3 e^{-x_0/\lambda}.
\end{equation}
Using the critical velocity parameter $v_c$ and eliminating
$x_0$, we find
\begin{equation} \label{phiresult}
	\phi = \frac{4\pi G\rho m}{\hbar v_0} \alpha \lambda^3 
		 \exp{\left[-\frac{2\beta}{\lambda}\left(\frac{v_c}{v_0}\right)^{1/2}\right]}.
\end{equation}
Below we compare this result to a detailed numerical model and find it to be reasonably accurate.

Note that the model assumes that the quantum reflection
occurs only from the Casimir-Polder potential. If the incident velocity is
too low, then the Yukawa potential itself can play a role and alter the behavior. This is avoided if
$|(1/kV) dV/dx| < 1$ for $x > x_0$, as in Eq.~\eqref{Ucondition}. For the Yukawa potential 
\eqref{yukawa}, $|(1/V) dV/dx| = 1/\lambda$, so we require $k_0 \lambda > 1$.
This sets a velocity limit 
\begin{equation} \label{vlimit}
v_0 > \frac{\hbar}{m\lambda} = \frac{4\beta}{\lambda} v_c.
\end{equation}
To be consistent with $v_0 \lesssim v_c$ we require $\lambda \gtrsim 4\beta$, which
limits the short-range behavior that can be probed. 

\begin{table}
	\begin{tabular}{llll}
		Species \hspace{1em} & $\beta$~(\textmu m) \hspace{1em} & $v_c$~(mm/s) \hspace{1em}&  $\phi/\alpha$ \\ \hline
		$^7$Li & 0.25 & 9.3  & $1.7\times 10^{-9}$ \rule{0ex}{3ex} \\
		$^{23}$Na & 0.46 & 1.5  & $1.7\times 10^{-8}$ \\
		$^{41}$K & 0.87 & 0.44  & $5.0\times 10^{-8}$ \\
		$^{87}$Rb & 1.4 & 0.13  & $2.1\times 10^{-7}$ \\
		$^{133}$Cs & 1.9 & 0.06 & $4.5\times 10^{-7}$
	\end{tabular}
	\caption{Quantum reflection parameters for different alkali atoms from a gold surface. 
		The $\beta$ values
		for K, Rb, and Cs are derived from the experimental measurements of Shih and Parsegian \protect\cite{Shih1975}. The $C_3$ values reported there are a factor of 3.5 lower than
		theoretical values for a perfect conductor that are reported in Ref.~\protect\cite{Derevianko1998}. To obtain $\beta$ values for Li and Na, the perfect-conductor values were scaled down by the same factor. 
		The critical reflection velocity $v_c = \hbar/4m\beta$ gives the largest 
		incident velocity for which quantum reflection will be efficient. 
		The phase sensitivity $\phi/\alpha$ is evaluated at $\lambda = 10$~\textmu m, using  Eq.~\protect\eqref{phiresult} with $v_0/v_c = 4\beta/\lambda$
    	and a material density $\rho = 19$ g/cm$^3$. \label{tab1} }
\end{table}

Table~\ref{tab1} gives values of $\beta$, $v_c$ and $\phi/\alpha$ for the alkali 
elements on a conducting surface. We see that the highest sensitivity is achieved with the heaviest 
species.
However, the low $v_c$ values are potentially challenging. The corresponding kinetic energies are
90 pK for Rb and 30~pK for Cs, both comparable to the lowest temperature
experiments to date \cite{Kovachy2015,Gaaloul2022a}. We focus here on Rb due to its less
stringent cooling requirement as well as its simpler process for condensate production. 

Because of the low temperatures involved, it would be difficult to 
prepare an incident wave function with well-defined energies at this level.
We therefore consider an alternative
approach wherein the confining forces on a stationary trapped cloud are turned off and the cloud
is allowed to expand into the surface. In this situation there is a spread of  
incident velocities, but the expansion velocity can be reasonably well-defined and can be calculated in a
mean-field model. As seen in Section III, the numerical
results for this case remain in general agreement with the analytical approximation.

\section{Numerical Modeling}

Numerical modeling of quantum reflection is challenging because the short-range atom-surface
interactions are complicated. One way to avoid dealing with this complexity is to introduce
an absorbing boundary condition in front of the surface, but it can be difficult to ensure
the boundary is perfectly absorbing \cite{Antoine2008}. 
Any spurious reflections can distort the desired quantum
reflection process. The alternative approach we use here is to smoothly transition the 
attractive Casimir-Polder potential to a constant value and allow atoms to continue
propagating toward negative $x$ \cite{Galiffi2017}. The resulting potential is
\begin{equation} \label{Ubar}
	\bar{U}(x) = \begin{cases}
	\displaystyle	-\frac{C_4}{(x+b)x^3} & (x > x_2) \\
		U_0 + A(x-x_1)^2 & (x_1 < x < x_2) \\
	\rule{0pt}{3ex}	U_0 & (x < x_1)
		\end{cases}
\end{equation}
where $x_1$ and $x_2$ define the boundaries of a transition zone. We abbreviate $b = 3\lambda_a/2\pi^2$. The potential and its first derivatives are continuous if 
\begin{equation}
	A = \frac{C_4}{2} \frac{4x_2 + 3b}{x_2^4(x_2+b)^2(x_2-x_1)}
\end{equation}
and
\begin{equation}
	U_0 = -\frac{C_4}{(x_2+b)x_2^3} - A(x_2-x_1)^2.
\end{equation}
A similar modification is applied to the Yukawa potential to produce a modified perturbation
 $\bar{V}(x)$.

We require the outer edge of the transition zone, $x_2$, to be well below the
quantum reflection point $x_0$ of Eq.~\eqref{xm}. For Rb atoms with $v \approx v_c$, 
we have $x_0 \approx 3~\mu$m. The transition zone width $x_2 - x_1$ must
be sufficiently large to provide a smooth potential and avoid spurious reflections. However,
using a very wide transition zone leads to a large value of $|U_0|$ and a high atomic 
velocity in the $x<x_1$ region,
which then requires an impractically small grid spacing for the numerical solution.  For
the calculations presented here, we use
$x_1 = 0.4$~\textmu m and $x_2 = 0.5$~\textmu m, giving $U_0/C_4 = -18$~\textmu m$^{-4}$. 
We verified that the numerical results do not vary significantly as long as $x_1$ and $x_2$
remain small compared to $x_0$, and $x_2-x_1$ is larger than about 0.1~\textmu m. 
We found that a grid spacing of 10~nm was sufficient to
give a phase accuracy better than $10^{-2}$~rad for all parameters considered here.

We first consider the case of non-interacting atoms incident from infinity with a 
well-defined velocity $-v_0$. 
A fraction of the atoms will reflect to form an interference pattern, while the remainder
continue propagating towards $x = -\infty$ with velocity $v_t = -(v_0^2-2U_0/m)^{1/2}$. 
Here we solve the time-independent Schr\"odinger equation
\begin{equation}
-\frac{\hbar^2}{2m} \frac{\partial^2 \psi}{\partial x^2} + \big[\bar{U}(x)+\bar{V}(x)\big]\psi = E_0\psi
\end{equation}
with $E_0 = mv_0^2/2$. Since we expect only a single wave
for $x\leq 0$, we start the solution at $x = 0$ 
with $\psi(0) = 1$ and $\psi'(0) = -ik_t \equiv -imv_t/\hbar$.
We integrate towards positive $x$ 
to find the solution for $x$ large enough that $|\bar{U}(x)|\ll E_0$.  
In this large-$x$ region, we expect $\psi(x)$ to be of the form $Ae^{-ik x} + Be^{ik x}$.  The interference pattern then has the form $|\psi|^2 = |A|^2 +|B|^2 +2|A||B|\cos(2kx + \theta)$ where $\theta$ is the phase difference between $A$ and $B$.  To evaluate $\theta$, we 
numerically find a location $x_\mathrm{m}$ of an extremum in $|\psi|^2$. We analytically evaluate $k(x_\mathrm{m})$ using \eqref{kofx}, and then
take $\theta = -2k x_\mathrm{m}$, modulo $2\pi$. 
We repeat the calculation with $\alpha = 0$, and compute the Yukawa phase $\phi$ by subtracting the $\theta$ phase calculated using the nearest extremum.
The solid points in Fig.~1 show the results, and agree quite well with Eq.~(10) for velocities above the velocity limit
from Eq.~\eqref{vlimit}. 

\begin{figure}
	\includegraphics[width=3.3in]{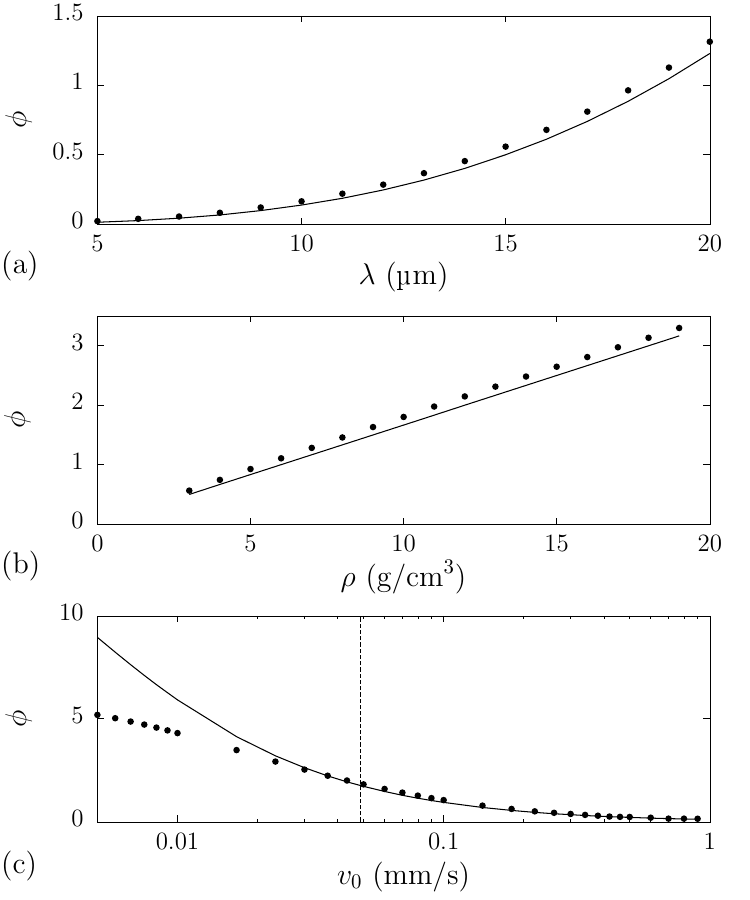}
	\caption{Comparison between numerical results for a non-interacting gas (data points) and the analytical model of Eq.~\protect\eqref{phiresult} (curves).
		(a) The Yukawa perturbation phase $\phi$ as a function of the Yukawa length $\lambda$, 
		for $^{87}$Rb atoms with incident speed $v_0 = 0.2$~mm/s and a perturber density
		$\rho = 3$~g/cm$^3$. (b) Phase $\phi$ vs.~perturber density $\rho$, for $v_0 = 0.2$~mm/s
		and $\lambda = 15$~\textmu m. (c) Phase $\phi$ vs.~incident speed $v_0$, for 
		$\lambda = 15$~\textmu m and $\rho = 3$~g/cm$^3$. The vertical dashed line shows the
		speed limit of Eq.~\protect\eqref{vlimit}. For all curves, $\alpha = 10^7$.
	}
\end{figure}

To model the alternative experiment in which a stationary atom cloud is allowed to 
expand into a surface, we solved the time-dependent problem. Here we allow for atomic interactions
in the one-dimensional Gross-Pitaevskii (GP) equation \cite{Dalfovo1999},
\begin{equation} \label{GPE}
i\hbar \frac{\partial \psi}{\partial t} = -\frac{\hbar^2}{2m} \frac{\partial^2 \psi}{\partial x^2}
+ \big[\bar{U}(x)+\bar V(x)\big]\psi + g|\psi|^2\psi,
\end{equation}
with
\begin{equation}
g = \frac{4\pi \hbar^2 a}{m} \sigma,
\end{equation}
where $a=5.05$~nm is the three-dimensional scattering length and $\sigma = N/A$ is the area particle density
for $N$ atoms in transverse area $A$. 
Pasquini {\em et al.}~\cite{Pasquini2006} found that a mean-field calculation yielded good agreement
with observed reflection probabilities for their experiments at much stronger interactions than
we consider, so we expect the model to be adequate here.

In our calculation, the initial wave function is first determined by solving
the GP equation in imaginary time for atoms confined in a harmonic trap potential
\begin{equation}
	V_{\rm trap}(x) = \frac{1}{2} m\omega^2 (x-x_t)^2,
\end{equation}
with $\omega = 2\pi\times 0.5$~Hz and $x_t = 150$~\textmu m.
We then turn off the potential and allow the atoms to expand in real time
towards the surface at $x=0$.
We solve Eq.~\eqref{GPE} using a Crank-Nicolson scheme with a grid spacing of $10$~nm
and a time step of $2.5$ \textmu s \cite{Bao2003}.
Figure 2 shows the atomic density distribution that results after an evolution time of
1.1~s, with and without the perturbation potential $\bar V$. The oscillations 
are the interference effect, and the phase shift from the perturbation is clear.

\begin{figure}
	\includegraphics[width=3.3in]{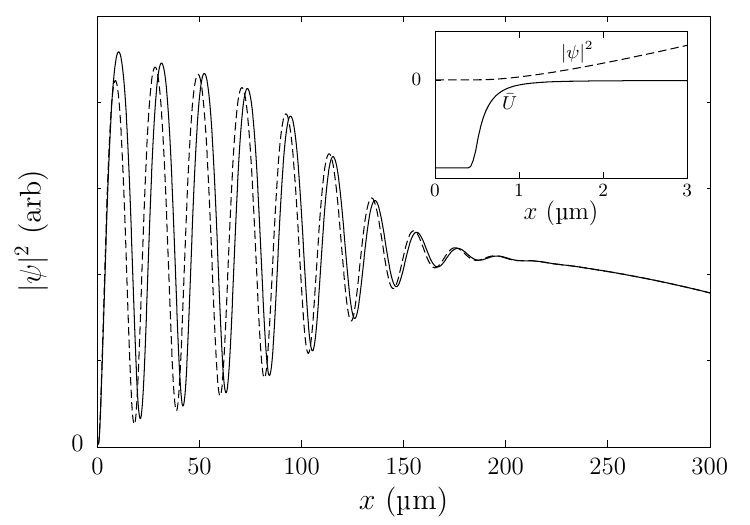}
	\caption{Numerical solutions for quantum reflection of an interacting $^{87}$Rb gas, using the 
		Gross-Pitaevskii equation \protect\eqref{GPE}.
		Main plot: Atom density $|\psi|^2$ after reflection from a gold surface.
	The oscillations show the interference between the incident and reflected
	wave functions. The solid curve is obtained with only the Casimir-Polder surface
	interaction, while the dashed curved is obtained when including a Yukawa potential
	with $\alpha = 10^7$ and $\lambda = 15$~\textmu m. For both curves, we use an areal density
	$\sigma = 10^{10}$ atoms/cm$^2$. Inset: Detail showing the atomic density (dashed)
	and the truncated surface interaction $\bar{U}$ (solid) at small distances.}
\end{figure}

We again extract the Yukawa phase $\phi$ by comparing the two interference patterns, with
results shown in Fig.~3. Here the data points correspond to different values of $\alpha$
and $\rho$, with the phase normalized to the values used for the solid circles. As seen,
the different results are compatible with each other when normalized this way. 

The curve in Fig.~3 is the analytical result from Eq.~\eqref{phiresult}. 
However, in this configuration there is no definite incident velocity $v_0$.
Instead, we estimate an effective $v_0$ using the spatial period $\Lambda$ of the interference
fringe observed in the numerical solution at the position where the phase is measured. 
The pattern here is produced by atom waves with two different velocities $v_{\rm inc}$ and
$v_{\rm ref}$, where we expect $|v_{\rm ref}|\gg |v_{\rm inc}|$, since the reflected atoms have
traveled considerably further than the incident atoms. For simplicity, we neglect
$v_{\rm inc}$ and set $|v_{\rm ref}| = 2\pi\hbar/m\Lambda$. We then use
this velocity for $v_0$ in Eq.~\eqref{phiresult}. We find that $\Lambda$ does depend 
slightly on the Yukawa perturbation, but again for simplicity, in Fig.~3 we use 
a fixed value of $v_0$ determined from the interference pattern with no perturbation. 
As seen, this yields reasonably good agreement with the numerical solutions.

\begin{figure}
	\includegraphics[width=3.3in]{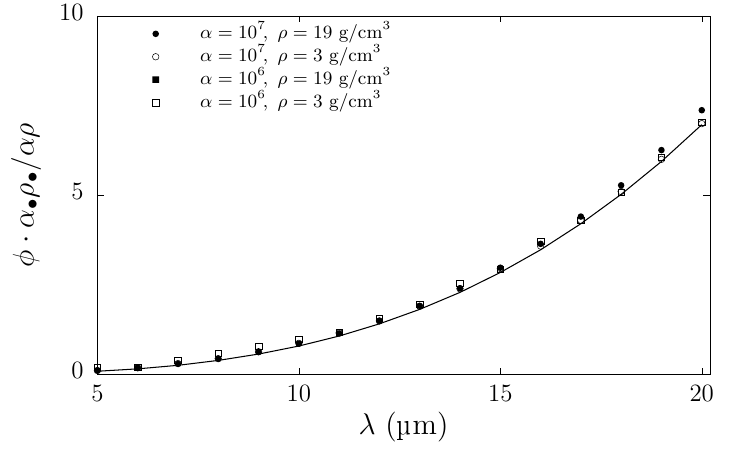}
	\caption{Numerical and analytical results for an interacting $^{87}$Rb gas reflecting
		from a gold surface. The Yukawa phase $\phi$ is shown as a function of the
		interaction length $\lambda$ for various values of interaction strength $\alpha$ and
		surface density $\rho$, as indicated. To illustrate the scaling behavior, 
		the phase values are multiplied by $\alpha_\bullet \rho_\bullet/\alpha\rho$, where
		$\alpha_\bullet = 10^7$ and $\rho_\bullet = 19$~g/cm$^3$. 
		Data points are numerical results from
		the Gross-Pitaevskii equation, and the curve is the analytical result of 
		Eq.~\protect\eqref{phiresult}. The analytical formula was evaluated using
		$v_0 = 0.225$~mm/s, which was obtained from the observed period of the interference pattern
		as seen in Fig.~2.}
\end{figure}   

We also used the time-dependent model to investigate the effect of atomic interactions
on the interference phase. Unsurprisingly, this effect is significant.
Here we set the Yukawa interaction to zero and instead observe how the interference
phase varies with the atom density $\sigma$. For the conditions of Fig.~2, we find that
a 10\% variation in $\sigma$ leads to a phase shift of about 1 rad. This matters because
in an experiment, the atom number typically fluctuates from run to run, making the
interactions a source of phase noise. One way to mitigate this effect is to operate
at a Feshbach resonance where interactions can be suppressed. Alternatively, the number
of atoms can be determined with good accuracy at the same time that the interference pattern
is observed. By comparing runs having similar number values, the impact of the Yukawa interaction
could be extracted even in the presence of noise.

Two possible explanations for the atom-number sensitivity can be considered. First,
the chemical potential of the initially-trapped atoms depends on $\sigma$, which in turn affects
the expansion velocity of the released atoms and thus the reflection phase. Second, interactions that occur during 
the reflection process could modify the phase directly. To isolate these possibilities, we
observe how the wavelength of the interference pattern varies with $\sigma$, and interpret
that in terms of an effective velocity variation. We then use the non-interacting model to
see how this velocity variation would affect the phase. We find that the interaction-induced velocity variation
corresponds to a phase shift of only a few mrad, 
much smaller than observed in the model. We conclude
that the phase sensitivity comes primarily 
from non-trivial interaction effects arising during the reflection process.

Both models considered here are one-dimensional, for simplicity. In contrast, the experimental
realizations of Ref.~\cite{Pasquini2006} showed 
significant transverse excitations of the condensate, 
which were attributed to mean-field interactions between the incident and reflected waves. 
In that work, the chemical potential of the condensate, $\mu$,
was about thirty times larger than the transverse confinement energy $\hbar\omega_\perp$, so interactions played a 
dominant role in the transverse dynamics. In comparison, we propose using much lower atomic densities,
such that the interaction forces are small. The conditions of Fig.~2 give a chemical potential $\mu$ of about $2\pi\hbar\times 3$~Hz
after the atoms have expanded into the surface.
These conditions can be achieved with $10^4$ atoms in a trap with
$\omega_\perp \approx 2\pi\times 5$~Hz, such that $\mu/\hbar\omega_\perp < 1$ and the atoms are entering the one-dimensional
regime \cite{Goerlitz2001}.
Further, the time scale for the interference pattern to form is approximately
$2\pi/k_0v_0 = 2\pi\hbar/mv^2$, about 0.1~s at $v_0 = 0.2$~mm/s. Since this is less than one period of the transverse motion,
excitations that do occur will not have time to produce significant distortions. Nonetheless, it will be useful
to implement a full three-dimensional GP model and investigate transverse effects in more detail.

\section{Implementation}

We consider here practical aspects of the proposed measurements and estimate the sensitivity 
that could be achieved. We take the atomic source to be a Bose-Einstein condensate
confined in trap that is positioned near the reflecting wall. For instance,
an Ioffe-Pritchard magnetic trap could be aligned with its weak axis perpendicular to the wall.
We envision a condensate
being prepared in this potential, moved near to the wall, and then released by turning off the
$x$ confinement. Once the interference pattern is established, it can be observed using standard
absorption imaging. 
To probe for short-range gravitational effects, it is necessary to vary the density
of the reflecting test mass. It is critical to do so without affecting the surface position, the Casimir-Polder potential, and any electrostatic patch potentials. 
An established way to achieve this is to screen the 
test mass with a thin conducting membrane \cite{Geraci2010}. The membrane thickness should be 
small compared to the Yukawa length $\lambda$, here 5~\textmu m or larger.
The membrane must also be thick enough to effectively screen the variable surface:
Bennett and O'Dell have calculated that 50~nm is sufficient for a good conductor like gold
\cite{Bennett2019}.
We therefore propose a thickness of order 0.5~\textmu m, which satisfies both criteria well.
Suitable membrane materials include silicon nitride or beryllium, which can be as thin as
a few nm and subsequently coated to the thickness desired \cite{Dwyer2017,Tanimura2024}.
As long as the surface is smooth on the scale of the thickness, roughness
effects will be negligible.

The test mass itself could be implemented as a uniform material whose position
can be adjusted to be either near (distance small compared to $\lambda$) or far
(distance large compared to $\lambda$) from the membrane. Alternatively, the test mass
could be constructed by bonding two materials of low and high density, which is then
moved parallel to the membrane in order to vary the density near the atoms. A potential
advantage of the second technique is that any electrostatic coupling between the test mass and 
membrane would remain approximately constant, suppressing induced effects on 
the membrane. Suitable contrasting materials include gold
($\rho = 19$~g/cm$^3$) and glass ($\rho = 3$~g/cm$^3$). 
A thin coating of gold over the glass material would further stabilize coupling to the 
membrane by presenting a uniform surface interaction. Figure 4 illustrates a possible configuration
for this approach.

\begin{figure}
	\includegraphics[width=3.3in]{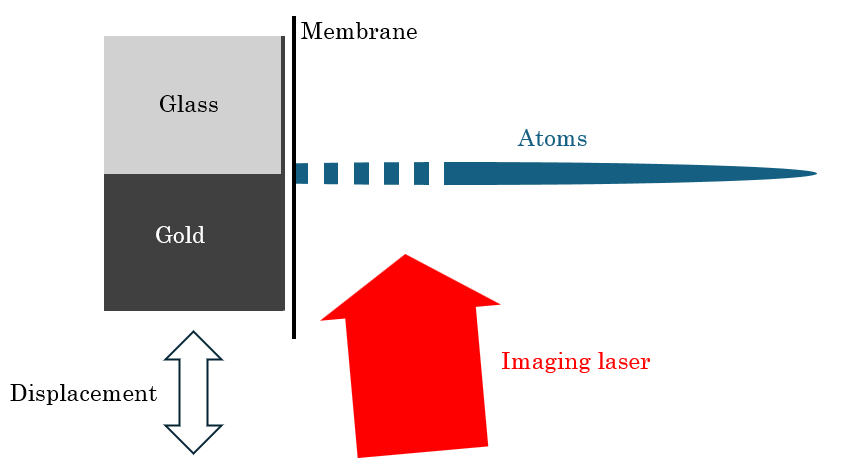}
	\caption{(Color online) 
		Schematic layout for a quantum reflection interferometer. A Bose condensate
		(blue) in an elongated trap potential is allowed to expand into a conductive
		membrane. The membrane sits just in front of a test mass consisting of bonded
		gold and glass blocks. The test mass can be translated roughly 50~\textmu m
		parallel to the surface to vary the density with which the atoms interact.
		An absorption imaging laser beam (red) passes through the atoms at a small angle,
		reflects from the membrane, and is then imaged by a camera (not shown). A secondary
		imaging system perpendicular to the plane of the drawing is used to locate
		the atoms relative to the boundary between test-mass sections.}
\end{figure}

For imaging, a resonant laser beam can pass through the atoms, parallel to the membrane.
With a fringe spacing from 5~\textmu m to 50~\textmu m, high spatial resolution is not required.
However, imaging atoms near a surface can be challenging due to diffraction from the
surface edge. One solution is to take
advantage of the fact that the fringe spacing is comparable to the approximately $10$-\textmu m 
transverse width of the condensate, meaning that the fringes can still be
observed using an imaging beam that travels at a small angle relative to the surface and
subsequently reflects off of the membrane, thus avoiding the edges.
 A similar method was used by Harber {\em et al.}\ 
to image atoms as close as 6~\textmu m to a surface \cite{Harber2005}. 
This method would produce two images of the atoms, one direct and one reflected. 
The center of symmetry of the combined image could be used to determine the location
of the membrane itself and provide a phase reference. The trapping axis will need to be
aligned normal to the membrane with an angular accuracy better than $v_\perp/v_0$, where
$v_\perp = \sqrt{\hbar\omega_\perp/m} \approx 0.15$~mm/s gives the transverse velocity. 
Even at $v_0 \approx 1$~mm/s, the precision is about five degrees which is not challenging.

The sensitivity of the interferometer will depend on how accurately the fringe phase can be 
determined, which is likely to be limited by the interaction noise described above.
We suppose here that the atom number can be measured with 1\% accuracy, 
corresponding to phase noise of 0.1~rad. By averaging on the order of $10^4$ measurements,
the sensitivity can be reduced to about 1 mrad, which we take as a practical limit.
Assuming an experimental rate of 250 runs per day, this could be achieved in a
two-month measurement campaign. Any other noise sources that fluctuate from run to run 
can also be averaged over, 
so they will not impact the sensitivity so long as the variations are small
compared to those from the interaction effect.

In contrast, slowly varying noise effects are excluded by 
the differential nature of the proposed technique. 
For instance, the interaction between the atoms and the membrane will be modified
by the presence of surface contaminants, which can be expected to vary over time. 
An obvious contaminant is rubidium.
Based on the electric field measurements of Ref.~\cite{Obrecht2007b},
we estimate that adding one monolayer of Rb atoms to the membrane
would shift the $C_4$ coefficient by about 30\%, which according to our 
models would cause a shift in the
the reflection phase of about 0.1~rad. This is significant compared to our proposed
precision, but as long at it is reasonably constant on the time scale of a single measurement run,
it will cancel out when the two source-mass configurations are compared. Since the time scale for Rb
coverage to vary is $10^6$~s at room temperature \cite{Obrecht2007b}, we do not expect this to be a significant source of noise.

The Yukawa sensitivity depends on the incident velocity $v_0$, which can be controlled
through the initial confinement strength of the atom trap along $x$ and the timing
of the imaging pulse. Larger velocities can be obtained by releasing the atoms
from a moving trap. The optimum sensitivity for a given $\lambda$ occurs at 
$v_{\rm opt} = (\beta^2/\lambda^2) v_c$, but achieving this is limited on the low side
by accessible atom energies, and on the high side by the critical velocity for
quantum reflection. We estimate that velocities in the range of 50~\textmu m/s
to 1~mm/s can be used for measurement.

\begin{figure}[t!]
	\includegraphics[width=3.4in]{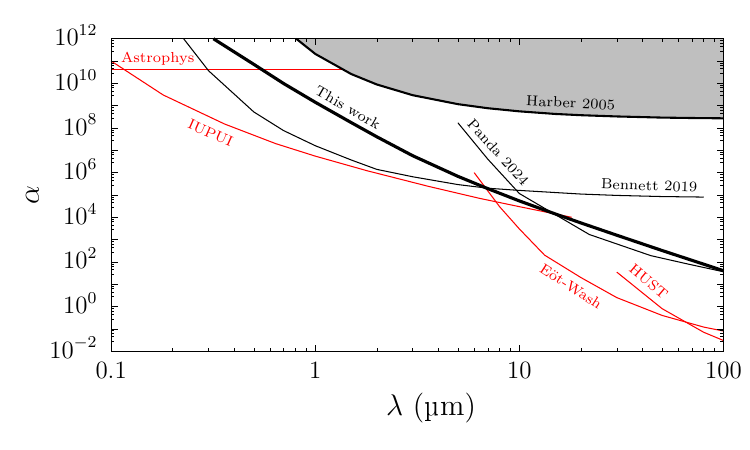}
	\caption{Constraints on the Yukawa $\alpha$ and $\lambda$ parameters from
		existing atomic (solid gray), existing macroscopic (red) and proposed atomic (black) measurements. 
		The Harber 2005 results \protect\cite{Harber2005}
		are measurements based on the oscillation frequency of atoms trapped near a surface. The
		Bennett 2019 \protect\cite{Bennett2019} curve is based on a proposed experiment using Bloch oscillations
		of trapped atoms near a surface. The Panda 2024 \protect\cite{Panda2024} curve is based on a proposed
		measurement using a lattice-based interferometer near a surface. The IUPUI \protect\cite{Chen2016},
		E\"ot-Wash \protect\cite{Lee2020}, HUST \protect\cite{Tan2020} and Astrophys \protect\cite{Fiorillo2025}
		curves are macroscopic results using micro-cantilevers, torsion pendulums, and neutron star observations.}
\end{figure}

With these assumptions, we achieve the sensitivity limits shown in Fig.~5. Here the gray
shading indicates existing limits for microscopic particles. The heavy black curve shows our result, and the
lighter black curves are for two other proposed
microscopic measurements. The red curves show existing limits for macroscopic measurements.  Given the assumption of a mass-dependent force, it is difficult for atomic measurements to reach the sensitivity 
of a macroscopic object. However, the method proposed here would improve on existing
atomic limits by several orders of magnitude, providing a useful constraint
on chameleon-style theories with short-range interactions. 

At lower $\lambda$ values, the sensitivity determined
here is comparable to that calculated by Bennett and O'Dell \cite{Bennett2019}, 
but potentially easier to implement experimentally
since a high-precision Bloch oscillation measurement is not required.
At higher-$\lambda$ values, our approach is competitive with proposed improvements to an optical-lattice 
interferometer experiment by Panda {\em et al.} \cite{Panda2024}, but may again be
easier to implement.

\section{Conclusions}

Our results illustrate the possibility of using a quantum reflection
interferometer to probe surface effects. Applied, as here, to searching for an 
anomalous short-range force, we find that the method is promising for improving
existing limits on coupling to microscopic bodies like atoms. 

The method could more broadly be applied to the measurement of ordinary atom-surface
interactions, such as electrostatic patch potentials or the Casimir-Polder force itself.
Absolute measurements may be challenging due to the significant phase contribution
from atomic interactions, but in many cases differential measurements are still useful.
For instance, it should be possible to measure the temperature dependence of the Casimir-Polder
force by varying the temperature of the surface \cite{Obrecht2007}, 
or to investigate slow temporal variations
in patch potential effects \cite{Yin2014}. 
Alternatively, using a Feshbach measurement to suppress
interactions might allow for precision measurements of the $C_4$ 
coefficient.

Applications such as these are likely to benefit from the fact that the simple
analytical model presented here agrees well with numerical calculations. The model
can be readily adapted to other forms for the perturbation potential $V$.

Interferometric methods such as this are most easily analyzed in terms of potential 
energies rather than forces, but it is nonetheless interesting to evaluate 
the force sensitivity implied by our results. For the conditions of Fig.~2, the 
nominal reflection distance is $x_0 \approx 2.3$~\textmu m, at which point the
strength of the Yukawa force $-dV/dx$ is quite small,
$1.5\times 10^{-28}$~N. As seen in the figure,
this force yields a distinguishable phase shift in a single measurement. The
interaction time with the surface can be estimated as $\lambda/v_0 \approx 75$~ms,
using $v_0 \approx 0.2$~mm/s. This 
illustrates that the method can provide high force sensitivity in a small volume and short time,
which could be useful for a variety of applications.

\begin{acknowledgments}

The authors gratefully acknowledge advice from Mark Edwards and Edoardo Vitagliano,
and comments on the manuscript from Christian Brandt, Zekun Chu, and Itzal De Urioste Terrazas.
This work was supported by the National Science Foundation, Grant No. 2110471. 
J.R.B is supported by the NSF Graduate Research Fellowship Program under Grant No. DGE 2137420.
	
\end{acknowledgments}


\bibliographystyle{apsrev4-2}

\bibliography{qreflect}
	
\end{document}